\documentclass[eqsecnum,showkeys,showpacs,nofootinbib,aps]{revtex4}

\usepackage{graphicx}
\def\bea{\begin{eqnarray}}
\def\eea{\end{eqnarray}}

\newcommand{\nn}{\nonumber}
\newcommand{\na}{\nabla}
\def\beq{\begin{equation}}
\def\eeq{\end{equation}}

\def\pa{\partial}

\def\na{\nabla}
\newbox\pippobox

\def\btheta{\bar{\theta}}
\def\bpsi{\bar{\psi}}
\def\bbeta{\bar{\beta}}
\def\btheta{\bar{\theta}}
\def\bdelta{\bar{\delta}}

\def\Dt{D_{t}}

\begin{document}
\title{Isospin particle on $S^{2}$ with arbitrary number of supersymmetries}
\author{Soon-Tae Hong}
\email{soonhong@ewha.ac.kr} \affiliation{Department of Science
Education and Research Institute for Basic Sciences,\\
Ewha Womans University, Seoul 120-750 Korea}
\author{Joohan Lee}
\email{joohan@kerr.uos.ac.kr} \affiliation{Department of Physics,
University of Seoul, Seoul 130-743 Korea}
\author{Tae Hoon Lee}\email{thlee@ssu.ac.kr}
\affiliation{Department of Physics and Institute of Natural Sciences,\\
Soongsil University, Seoul 156-743 Korea}
\author{Phillial Oh}
\email{ploh@newton.skku.ac.kr} \affiliation{Department of Physics
and Institute of Basic Science,\\ Sungkyunkwan University, Suwon
440-746 Korea}
\date{\today}%
\begin{abstract}
We study the supersymmetric quantum mechanics of an isospin particle
in the background of spherically symmetric Yang-Mills gauge field.
We show that on $S^{2}$ the number of supersymmetries can be made
arbitrarily large for a specific choice of the spherically symmetric
$SU(2)$ gauge field. However, the symmetry algebra containing the
supercharges becomes nonlinear if the number of fermions is greater
than two. We present the exact energy spectra and eigenfunctions,
which can be written as the product of monopole harmonics and a
certain isospin state. We also find that the supersymmetry is
spontaneously broken if the number of supersymmetries is even.
\end{abstract}
 \preprint{hep-th/yymmnn} \maketitle


The study of supersymmetric quantum mechanics of charged particle in
the background of magnetic monopole \cite{cole} revealed several
interesting aspects the system \cite{vine,deJonghe, Linden,hllo06}.
Especially, the system is known to possess a hidden supersymmetry
commuting with the original supersymmetry which squares to the
Hamiltonian of the system. This feature is present both in the case
\cite{deJonghe} of a charged particle in the background of the Dirac
monopole and the case \cite{Linden} of an isospin particle in the
Wu-Yang monopole \cite{wuya} background. In Ref. \cite{deJonghe}, it
was shown that the hidden supersymmetry can be identified with the
usual supersymmetry of a charged particle if it is restricted on a
sphere. This led to the manifest extended supersymmetric formulation
of charged particle on $S^2$ in the Dirac magnetic monopole
background. In Ref. \cite{hllo06} the complete energy spectra and
the corresponding eigenfunctions were obtained and the issue of the
invariance of the ground energy states under the supersymmetry
transformation were discussed, and it was shown that spontaneous
breaking of supersymmetry occurred for certain values of the
monopole charge. Furthermore, it was shown that the system on
$S^{2}$ admits $N=4$ real supersymmetries in contrast to the $N=1$
and $N=2$ supersymmetries on $R^3$ \cite{vine}. It turns out that
the supersymmetry generators in that case form $SU(2\vert 1)$
\cite{freu} superalgebra rather than the standard supersymmetry
algebra in which the superchages square to the Hamiltonian of the
system. In this context it would be interesting to add more
fermionic degrees of freedom and check the consequent superalgebra.

In this paper, we study the supersymmetric quantum mechanics of an
isospin particle on $S^2$ in the background of the Wu-Yang monopole
\cite{wuya}, a spherically symmetric solution to the sourceless
$SU(2)$ Yang-Mills equations. In particular, we address the issue of
whether the number of internal supersymmetries can be enlarged. The
action principle for the supersymmetric isospin particles was given
in Refs.~\cite{bala77} and \cite{Linden}, which generalized bosonic
Wong's equation~\cite{wong}. This action has the properties that the
system is invariant under the simultaneous rotations of isospace and
ordinary space and under the $N=1$ real supersymmetry
transformation. We first formulate the supersymmetric isospin
particle system in the $CP(1)$ model approach~\cite{hllo06}. It is
shown that by introducing ${\cal N}$ complex fermion degrees of
freedom with a quartic fermion interaction term and considering the
system on a sphere in the background of Wu-Yang monopole gauge, the
number of complex supersymmetries ${\cal N}$ can be made arbitrary.
With an appropriate choice of ordering the quantized Hamiltonian can
be made invariant under the supersymmetry transformations. However,
the symmetry algebra including the supercharges becomes nonlinear
when the number of fermions is greater than two. We find that the
algebra becomes the usual supersymmetry algebra when ${\cal N}=1$.
If ${\cal N}=2$, the symmetry algebra can be identified with
$SU(2|1)$ superalgebra. For general ${\cal N}>2$ the algebra takes
the nonlinear form. We then consider the energy eigenvalue problem
and obtain the energy spectra and the eigenstates. It turns out that
the energy eigenfunctions are given by the product of monopole
harmonics \cite{wu} and certain isospin states. (See
Eq.~(\ref{isoeigen}) below.) By looking at the supersymmetric
structure of the ground state we find that the supersymmetry is
spontaneously broken depending on the number of ${\cal N}$: it is
unbroken when ${\cal N}$ is odd whereas half of the supersymmetries
are broken for even ${\cal N}$.

In order to proceed, let us consider an isospin particle in $R^3$ in
the background of the Wu-Yang monopole gauge field \cite{wuya},
$$A_{i}^{a}=-\epsilon^a{}_{ij}\frac{x^{j}}{gr^{2}},~~~A_{0}^{a}=0,$$
the field strength of which is given by $$
F_{ij}^{a}=\pa_{i}A_{j}^{a}-\pa_{j}A_{i}^{a}-g\epsilon^{abc}A_{i}^{b}A_{j}^{c}
=\epsilon_{ijk}\frac{x^{a}x_{k}}{gr^{4}}.$$ The $N=1$
supersymmetric Lagrangian \cite{bala77}, \cite{Linden} can be
written as \beq
L={1\over2}\dot{x}_i^2+{i\over2}\psi_i\dot\psi_i+i\btheta
\dot{\theta}-g\dot{x}_iA_i^aI^a -gS_iB_i^aI^a,\label{ordiL}\eeq
where $I^{a}=\frac{1}{2}\btheta\sigma^{a}\theta,$ and
$S_i:=-{i\over2}\epsilon_{ijk}\psi_j\psi_k$ denotes the spin. Its
Hamiltonian is given by $$ H= {1\over2}\dot{x}_i^2+gS_iB_i^aI^a=
{1\over2}(p_i-igA_i^aI^a)^2+gS_iB_i^aI^a,$$ and the total angular
momentum becomes \bea
J_i&=&\epsilon_{ijk}x_j(\dot{x}_k-gA_k^aI^a)+I_i+S_i\nn\\
&=&\epsilon_{ijk}x_j\dot{x}_k+ (\hat{x}_kI_k)\hat{x}_i
+S_i,\label{angular1}\eea where in the second line the explicit
form of the gauge field was used. This system is
known~\cite{deJonghe, Linden} to possess a hidden supersymmetry
commuting with the original supersymmetry. In
Ref.~\cite{deJonghe}, it is also shown that when the particle is
restricted to the sphere of constant radius they together form
$N=2$ supersymmetries. Therefore, in terms of the local
coordinates of $S^2$ one can write down a manifestly $N=2$
supersymmetric Lagrangian. In this paper, however, we will use
$CP(1)$ model type of approach expressing the Lagrangian in terms
of the $S^3$ variable $z=(z_1,z_2)$ satisfying $\bar{z}\cdot z=1$
and the complex spinor $\psi=(\psi_1,\psi_2)$ satisfying
$\bar{z}\cdot\psi=0$ and its conjugation relation, where we have
used the notation $A\cdot B= A_{1}B_{1}+A_{2}B_{2}$. The dynamics
on $S^2$ is recovered by imposing the local phase symmetry. The
Lagrangian can be written as \beq L=2|\Dt
z|^{2}+i\bpsi\cdot\Dt\psi+i\btheta\dot{\theta}-i(\bar{z}\sigma_{i}\Dt
z-\Dt\bar{z}\sigma_{i}z)I_i,\label{N1Lag}\eeq where $D_t$ defined
by $D_{t}z\equiv\dot{z}-iaz$ with $a$ given by $$
a=-\frac{i}{2}(\bar{z}\cdot\dot{z}-\dot{\bar{z}}\cdot
z)-\frac{1}{2}\bpsi\cdot\psi$$ is a covariant derivative with
respect to the $U(1)$ phase transformation. Consequently, the
Lagrangian describes a particle on unit $S^2$ although it is
written in $S^3$ variables. One can show that, using the unit
radial vector $x_{i}=\bar{z}\sigma_{i}z$ defining $S^2$, the last
term of Eq.~(\ref{N1Lag}) can be written as
$$\epsilon_{kij}\dot{x}_ix_jI_k+\bpsi\cdot\psi x_iI_i,$$ which is the
same as the interaction term of Eq.~(\ref{ordiL}) evaluated on the
unit $S^2$ in the background of the Wu-Yang monopole, if
$-(\bpsi\cdot{\psi})$ is identified as the classical analogue of
the radial component\footnote{For its quantum mechanical
definition, see Eq.~(\ref{sigma}). In fact, when restricted to
$S^2$, only the radial component of the spin survives.} of the
spin $S_i$. With the identification
\beq\psi_i={1\over\sqrt{2}}(\bar{z}\sigma_{i}\psi+\bpsi\sigma_{i}z),\label{fermid}\eeq
one can further show that the first two terms of the
Eq.~(\ref{N1Lag}) become the kinetic part of the isospin particle
on $S^2$. In this formalism, supercharges can be written as
$$Q=2\dot{\bar{z}}\cdot\psi,~~~~~~\bar{Q}=2\bpsi\cdot\dot{z}.$$ From these, two
real combinations can be obtained, one of which is the ordinary
supercharge restricted to the sphere and the other one is the
so-called hidden supercharge.

We now consider a many fermion generalization of this model. The
Lagrangian is given by \beq L=L_{1}+L_{2} \label{lag}\nn \eeq with
\bea L_{1}&=& 2|\Dt
z|^{2}+\frac{i}{2}(\bpsi_{\alpha}\cdot\Dt\psi_{\alpha}
-\Dt\bar{\psi}_{\alpha}\cdot\psi_{\alpha})
-\frac{1}{2}(\bar\psi_\alpha\cdot\psi_\alpha)^2,\nonumber\\
L_{2}&=&\frac{i}{2}(\btheta\dot{\theta}-\dot{\btheta}\theta)
-i(\bar{z}\sigma_{i}\Dt z-\Dt\bar{z}\sigma_{i}z)I_{i}, \nn\eea
where the fermionic variable
$\psi_\alpha=(\psi_{\alpha1},\psi_{\alpha2})$ has additional
flavor index $\alpha=1, \cdots ,{\cal N}$ and satisfies \beq
\bar{z}\cdot\psi_{\alpha}=\bar\psi_{\alpha}\cdot z=
0,\label{consts}.\eeq Covariant derivative $D_t$ is defined, as
before, by $D_{t}z\equiv\dot{z}-iaz$, but $a$ given by \beq
a=-\frac{i}{2}(\bar{z}\cdot\dot{z}-\dot{\bar{z}}\cdot
z)-\frac{1}{2}\bpsi_{\alpha}\cdot\psi_{\alpha}.\label{a}\eeq
Besides the trivial summation over the flavor indices this
Lagrangian differs from the previous one by quartic fermionic
interaction term.

A straightforward calculation shows that $L_1$ and $L_2$ are
separately invariant and the constraints are preserved under the
following supersymmetric transformations:
$$
\begin{array}{lll}
\vspace{0.2cm} \delta_{\alpha} z=\psi_{\alpha}, &\delta_{\alpha}
\bar{z}=0,
&\delta_{\alpha}\psi_{\beta}=0,\\
\vspace{0.2cm}
\delta_{\alpha}\bpsi_{\beta}=2i\na_{\alpha\beta}\bar{z},
&\delta_{\alpha}\theta=\frac{1}{2}(\bar{z}\sigma_{a}\psi_{\alpha})\sigma_{a}\theta,
&\delta_{\alpha}\btheta=-\frac{1}{2}(\bar{z}\sigma_{a}\psi_{\alpha})\bar{\theta}\sigma_{a},\\
\vspace{0.2cm}
\end{array}
$$ and
$$
\begin{array}{lll}
\vspace{0.2cm} \bdelta_{\alpha} z=0, &\bdelta_{\alpha}
\bar{z}=\bpsi_{\alpha},
&\bdelta_{\alpha}\psi_{\beta}=2i\na_{\alpha\beta}z,\\
\vspace{0.2cm} \bdelta_{\alpha}\bpsi_{\beta}=0,
&\bdelta_{\alpha}\theta=-\frac{1}{2}(\bar{\psi}_{\alpha}\sigma_{a}z)\sigma_{a}\theta,
&\bdelta_{\alpha}\btheta=\frac{1}{2}(\bar{\psi}_{\alpha}\sigma_{a}z)\bar{\theta}\sigma_{a},
\end{array}
$$ where
$$
\na_{\alpha\beta}\bar{z}=\delta_{\alpha\beta}\Dt
\bar{z}-\frac{i}{2}(\bpsi_{\beta}\cdot\psi_{\alpha}
-\delta_{\alpha\beta}\bpsi_{\gamma}\cdot\psi_{\gamma})\bar{z}.
$$
One can show that $a$ defined in Eq.~(\ref{a}) is invariant, so
that the transformations commute with $D_t$. Furthermore, it turns
out that the constraints are also preserved.

The momentum conjugate to $z$ is given by
$$ p=2D_t\bar{z}+{i\over2}(\bar\psi_\alpha\cdot\psi_\alpha)\bar{z}
-i\bar{z}\sigma_i(1-z\bar{z})I_i.$$ For the consistency of the
time evolution with the constraint Eq.~(\ref{consts}), the
momentum should satisfy $$p\cdot z+\bar{z}\cdot\bar{p}=0.$$
Furthermore, the local $U(1)$ phase symmetry gives rise to the the
Gauss law constraint
$$-i(\bar{z}\cdot\bar{p}-p\cdot z)-\bar\psi_\alpha\cdot\psi_\alpha=0.$$
The Hamiltonian can be written as \bea
H&=&2|D_t\bar{z}|^2-\bpsi_\alpha\cdot\psi_\alpha(\bar{z}\sigma_iz)I_i\nn\\
&=&2|\dot{z}-z(\bar{z}\cdot\dot{z})|^2 +
{1\over2}(\bpsi_\alpha\cdot\psi_\alpha)^2-\bpsi_\alpha\cdot\psi_\alpha(\bar{z}\sigma_iz)I_i\nn\\
&=&{1\over2}|(\bar{p}-z(\bar{z}\cdot\bar{p})-i(1-z\bar{z})\sigma_izI_i|^2
+{1\over2}(\bpsi_\alpha\cdot\psi_\alpha)^2-\bpsi_\alpha\cdot\psi_\alpha(\bar{z}\sigma_iz)I_i.\nn
\eea

To quantize this system one needs to compute the Dirac brackets.
The result can be summarized as follows in the quantum commutator
version\footnote{In transition from Dirac brackets to quantum
commutator there may arise an ordering ambiguity as noted in
Ref.~\cite{hllo06}. However, it can be absorbed if the Gauss Law
constraint is appropriately ordered. So, here we choose a
particular ordering.}:\beq
\begin{array}{ll}
\vspace{0.2cm}
\left[p_{m},z_{n}\right]=-i\delta_{mn}+\frac{i}{2}\bar{z}_{m}z_{n},
&\left[p_{m},\bar{z}_{n}\right]=\frac{i}{2}\bar{z}_{m}\bar{z}_{n},\\
\vspace{0.2cm}
\left[p_{m},p_{n}\right]=\frac{i}{2}(p_{m}\bar{z}_{n}-p_{n}\bar{z}_{m}),
&\left[p_{m},\bar{p}_{n}\right]=\frac{i}{2}(p_{m}z_{n}-\bar{p}_{n}\bar{z}_{m})
-\bpsi_{\alpha m}\psi_{\alpha n}-{3\over2}(\delta_{nm}-z_n{\bar z}_m) ,\\
\vspace{0.2cm} \left[\bpsi_{\alpha m},\psi_{\beta
n}\right]=\delta_{\alpha\beta}(\delta_{nm}-z_n{\bar z}_m),
&\left[p_{m},\bpsi_{\alpha n}\right]=i\bpsi_{\alpha m}\bar{z}_{n},\\
\vspace{0.4cm} \left[I_i,I_j\right]=i\epsilon_{ijk}I_k,&
\end{array}\nn
\eeq where indices $m, n=(1,2)$ are explicitly written and the
square bracket between two fermionic operators should be
interpreted as the anticommutator.

In terms of the variables
$$
\begin{array}{ll}
\vspace{0.2cm} \beta_{\alpha}=\epsilon_{mn}z_{n}\psi_{\alpha m
}\equiv\epsilon z\psi_\alpha,&
\bbeta_{\alpha}=\bpsi_{\alpha m}\epsilon_{mn}\bar{z}_{n}\equiv\bpsi_{\alpha}\epsilon\bar{z}, \\
\vspace{0.2cm} B_{m}=p_{n}A_{nm},& \bar{B}_{m}=A_{mn}\bar{p}_{n},\\
\tilde{U}_B=-i(\bar{z}\cdot\bar{p}-p\cdot
z)+\bar\psi_\alpha\cdot\psi_\alpha,
\end{array}
$$
the above commutation relations can be simplified~\cite{hllo06}:
\beq
\begin{array}{ll}
\vspace{0.2cm} \left[B_{m},z_{n}\right]=-iA_{nm},
&\left[B_{m},\bar{z}_{n}\right]=0,\\
\vspace{0.2cm}
\left[B_{m},B_{n}\right]=-i(\bar{z}_{m}B_{n}-\bar{z}_{n}B_{m}),
&\left[\bar{B}_{m},B_{n}\right]=(\tilde{U}_B+1)A_{mn},\\
\vspace{0.2cm} \left[\tilde{U}_B, z_m\right]=z_m,
&\left[\tilde{U}_B, B_m\right]=-B_m,\\
\vspace{0.2cm}
\left[\bbeta_{\alpha},\beta_{\beta}\right]=\delta_{\alpha\beta},
&\left[I^{a},I^{b}\right]=i\epsilon_{abc}I^{c},
\end{array}\label{comm}
\eeq  where we have defined $A_{mn}\equiv\delta_{mn}-z_m\bar{z}_n$
for notational convenience. These should be supplemented by the
adjoint relations and all other omitted commutators vanish.
Besides these commutation relations the following should be
satisfied:
$$B\cdot z=0, ~~~~ \bar{z}\cdot\bar{B}=0,$$
as operator identities. The Gauss law constraints can be written
as \beq
C_G\equiv\tilde{U}_B-2\bbeta_{\alpha}\beta_{\alpha}+\alpha_G=0,
\label{gauss}\eeq which has to be imposed on the physical states.
The constant $\alpha_G$ is the ordering parameter that will be
fixed. [See Eq.~(\ref{ordering}).]

Conversely, original variables can be recovered as follows:
$$p=B-{i\over2}(\tilde{U}_B-\bbeta_{\alpha}\beta_{\alpha})\bar{z},~~~~
\psi_{\alpha}=\epsilon\bar{z}\beta_{\alpha}.$$ Comparing this
expression with the definition of the momentum yields the following
interpretation \bea
2\nabla_t\bar{z}&\equiv&2\dot{\bar{z}}(1-z\bar{z})=
B+ i\bar{z}\sigma_i(1-z\bar{z})I_i,\nn\\
2\nabla_tz&\equiv&2(1-z\bar{z})\dot{z}=\bar{B}-i(1-z\bar{z})\sigma_izI_i.\nn\eea
In quantizing the Hamiltonian there arises ordering ambiguity. One
can fix one particular ordering and determine the Hamiltonian by
adding the terms arising from changing the operator ordering using
other requirement. In our case we will impose the invariance under
the supersymmetric transformations. Therefore, we will tentatively
choose the following form of the Hamiltonian: \beq H=
2\nabla_t\bar{z}\cdot\nabla_tz
+{1\over2}(\bpsi_\alpha\cdot\psi_\alpha)^2-\bpsi_\alpha\cdot\psi_\alpha
I_r +C\bpsi_\alpha\cdot\psi_\alpha+DI_r, \label{qham}\eeq where
$I_r\equiv(\bar{z}\sigma_iz)I_i$ denotes the radial component of
the isospin and the constants $C, D$ are ordering parameters to be
fixed later.

The quantum angular momentum operator satisfying the correct
commutation relations turns out to be \bea J_i &=&
-{i\over2}(p\sigma_iz-\bar{z}\sigma_i\bar{p})-\bar{z}\sigma_iz
+{1\over2}\bar{\psi}_\alpha\sigma_i\psi_\alpha+I_i\nn \\
&=&-{i\over2}(B\sigma_iz-\bar{z}\sigma_i\bar{B}-2i\bar{z}\sigma_iz)
-{1\over2}\tilde{U}_B\bar{z}\sigma_iz+I_i\nn
\\ &=&\epsilon_{ijk}\bar{z}\sigma_jz(\nabla_t\bar{z}\sigma_kz+\bar{z}\sigma_k\nabla_tz)
-{1\over2}\tilde{U}_B\bar{z}\sigma_iz+(\bar{z}\sigma_kzI_k)\bar{z}\sigma_iz.\label{angular}\eea
The first expression is the same as the Noether charge except the
ordering term which is necessary to produce the correct angular
momentum commutation relations, and the third expression shows the
decomposition of the angular momentum into the orbital part and the
the rest. Comparing the final expression with Eq.~(\ref{angular1})
suggests identifying the second term as the spin. Indeed, using
Eq.~(\ref{fermid}) one can show that, when restricted to $S^2$, only
the radial component of the spin survives due to the constraint
(\ref{consts}). Furthermore, with the choice of ordering constant
\beq \alpha_G= {\cal N}\label{ordering}\eeq the Gauss law
constraint, Eq.~({\ref{gauss}) can be written as \beq
C_G=\tilde{U}_B+2\Sigma,\eeq where \beq \Sigma\equiv
-{1\over2}[\bbeta_\alpha,\beta_\alpha]\label{sigma}\eeq can be
identified as the radial component of the total spin.\footnote{In
this paper we define $\Sigma$ to be the outward radial component of
the spin operator, so it differs from that of Ref. \cite{hllo06} by
the sign.} This expression can be obtained by directly imposing the
constraints on the total spin $S_i$ and the eigenvalues range from
$-{{\cal N}\over2}$ to ${{\cal N}\over2}$.

Quantum mechanical supercharges $Q_\alpha$ and $\bar{Q}_\alpha$ can
be readily obtained from the classical Noether charge because they
have no ordering ambiguity,
$$Q_\alpha = p\cdot\psi_\alpha+i\bar{z}\sigma_i\psi_\alpha I_i,
~~~\bar{Q}_\alpha =
\bar{\psi}_\alpha\cdot\bar{p}-i\bar{\psi}_\alpha\sigma_izI_i.$$
One can further show that they can be rewritten in the following
suggestive form: \beq Q_\alpha=J_+\beta_\alpha, ~~~
\bar{Q}_\alpha= J_-\bar{\beta}_\alpha,\label{qqbar}\eeq where \bea
J_+&\equiv&
2\nabla_t\bar{z}\epsilon\bar{z}=i\bar{z}\sigma_i\epsilon\bar{z}J_i\nn\\
J_-&\equiv&2\epsilon z\nabla_t z=-i\epsilon z\sigma_i
zJ_i.\label{jpm}\eea Together with the radial component of the
angular momentum,\beq
J_{r}=-\frac{1}{2}\tilde{U}_{B}+I_{r}=-\frac{1}{2}C_G+(\Sigma+I_{r}),\label{norm}\eeq
$J_\pm$ satisfy the usual $SU(2)$ algebra relations: \beq [J_+,
~J_-]=2J_r,~~[J_r, ~J_+]= J_+, ~~[J_r, ~J_-]=-J_-.\label{su2} \eeq

For quantum Hamiltonian, first observe that \bea
{1\over2}J^2&=&{1\over4}(J_+J_-+J_-J_+)+{1\over2}J_r^2\nn\\
&=&{1\over2}J_+J_-+{1\over2}J_r^2-{1\over2}J_r\nn\\
&=&2\nabla_t\bar{z}\cdot\nabla_t z +{1\over2}\Sigma^2+ \Sigma I_r
+ {1\over2}I_r^2-{1\over2}\Sigma -{1\over2}I_r.\eea Without the
forth term this expression would take the same form as the
classical Hamiltonian Eq.~(\ref{qham}). Therefore, from now on we
will choose \bea H &\equiv&
{1\over2}J^2-{1\over2}I_r^2\nn\\&=&{1\over4}(J_+J_-+J_-J_+)+I_r\Sigma
+{1\over2}\Sigma^2 \label{ourham}\eea as our Hamiltonian. It is
supersymmetric and rotationally invariant, and corresponds to
quantizing the classical Hamiltonian with a specific choice of the
ordering parameters $C, D$ in Eq.~(\ref{qham}). The second term
represents the spin-isospin coupling and the third term may be
interpreted as the spin-spin coupling which becomes trivial when
${\cal N}=1$, but not in general.

From Eqs.~(\ref{sigma}),(\ref{qqbar}) and (\ref{su2}) we find \bea
[Q_\alpha, \bar Q_\beta]
&=&J_+J_-\delta_{\alpha\beta}-2J_r\bbeta_\beta\beta_\alpha\nn\\
&=&{1\over2}(J_+J_-+J_-J_+)\delta_{\alpha\beta}
-2J_r(\bbeta_\beta\beta_\alpha-{1\over2}\delta_{\alpha\beta})\nn\\
&=&\Big(2H-2({{\cal N}-1\over{\cal N}})I_r\Sigma-({{\cal
N}-2\over{\cal N}})\Sigma^2\Big)\delta_{\alpha\beta}-2(I_r+\Sigma)
S_{\beta\alpha},\label{algebra}\eea where
$S_{\beta\alpha}\equiv\bbeta_\beta\beta_\alpha-{1\over{\cal
N}}(\bbeta_\gamma\beta_\gamma)\delta_{\alpha\beta}$ denotes
$SU({\cal N})$ generator. Note that the trace part is not the
Hamiltonian. In fact, the trace part alone does not commute with the
supersymmetry generators. One can show that
$$H={1\over2{\cal N}}[Q_\alpha, \bar Q_\alpha]+({{\cal
N}-1\over{\cal N}})I_r\Sigma+({{\cal N}-2\over2{\cal N}})\Sigma^2.$$
The conserved quantities associated with the symmetries of the
system are (i) the total angular momentum $\vec{J}=\vec{K}+\vec{I}$
where $\vec{K}$ is the angular momentum associated with spatial
rotations of $z$'s and $\psi_{\alpha}$'s~\cite{hllo06} and $\vec{I}$
is the isospin associated with isospace rotations of $\theta$'s,
(ii) the spin $\Sigma=-\bbeta_{\alpha}\cdot\beta_{\alpha}+{{\cal
N}\over2}$ (or the fermion number
$N_{F}=\bbeta_{\alpha}\cdot\beta_{\alpha}$) associated with the
global $U(1)$ phase symmetry of the fermion, (iii) the internal
$SU({\cal N})$ charge, $S^A\equiv
S_{\alpha\beta}T_{\alpha\beta}=\bar\beta_\alpha T^A_{\alpha\beta}
\beta_\beta$, associated with the fermion flavor, where $T^A$ are
the traceless Hermitian ${\cal N}\times{\cal N}$ matrices satisfying
$[T^A,~T^B]=if^{ABC}T^C$ and normalized by ${\rm tr}(
T^AT^B)={1\over2}\delta^{AB}$, and (iv) the supersymmetric
generators $Q_{\alpha}=J_+\beta_\alpha$,
$\bar{Q}_{\alpha}=J_-\bbeta_\alpha.$

Although it is not the usual supersymmetry algebra, there are
several interesting aspects. If the number of fermion species is
one, i.e., ${\cal N}=1$, Eq.~(\ref{algebra}) simply becomes $[Q,
\bar Q]=2(H+{1\over8})$ which is the ordinary ${\cal N}=1$ SUSY
algebra. If ${\cal N}=2$, it can be shown that $\Sigma
S_{\beta\alpha}=0$ identically and the algebra reduces to \beq
[Q_\alpha, \bar Q_\beta]=
2H\delta_{\alpha\beta}-2I_r\Big(S_{\beta\alpha}+
{1\over2}\Sigma\delta_{\alpha\beta}\Big).\eeq This algebra can be
cast in the form which is identical to the $SU(2|1)$ algebra. For
general ${\cal N}$ the algebra is nonlinear.

The commutator algebra given in Eq.~(\ref{comm}) can be concretely
represented on the Hilbert space composed of certain functions on
$S^3$ as follows:
$$B_m=-iA_{nm}{\partial\over\partial z_n}+i\bar{z}_m,
~~~ \bar{B}_m=-iA_{mn}{\partial\over\partial \bar{z}_n},~~~
\tilde{U}_B=z_m{\partial\over\partial
z_m}-\bar{z}_m{\partial\over\partial\bar{z}_m}.$$ Operators $z$
and $\bar{z}$ act as multiplication, and the fermion operators and
the isospin can be represented as usual in terms of matrices. From
these, representations for other physical operators can be
obtained. In particular, we find \beq
J_{i}=\frac{1}{2}\left(\bar{z}\sigma_{i}\frac{\pa}{\pa\bar{z}}
-\frac{\pa}{\pa z}\sigma_{i}z\right)+I_i.\eeq

Let us study energy eigenfunctions of the Hamiltonian
(\ref{ourham}). Since $J_i$ is the sum of two commuting angular
momentum operators $K_i$ and $I_i$ and the isospin is easy to
handle, first consider the representation theory of $K_i$. Besides
$K^2$, $K_3$ there is another operator commuting with them, which is
the radial component $K_r$ of $K_i$. Let $\vert k, k_3, k_r\rangle$
be their simultaneous eigenstates. It can be shown that $k_r$
behaves exactly as $k_3$ relative to $k$, i.e., they are integer
spaced and for a given $k$, $-k\le k_r\le k$. In fact, as $k_3$ is
raised and lowered by $K_1\pm iK_2$, $k_r$ is raised and lowered by
$K_{\pm}$ defined as in Eq.~(\ref{jpm}). Thus, one can construct all
$\vert k, k_3, k_r\rangle$ starting from \beq\vert k, k_3=k,
k_r=k\rangle\sim \bar z_1^{2k}\eeq via a successive application of
the lowering operators $K_1-iK_2$ and $K_-$. For instance, one finds
\beq \vert k, k_3=k, k_r\rangle\sim z_2^{k-k_r} \bar z_1^{k+k_r}.
\label{monopoleharmonics} \eeq They are in fact monopole
harmonics.\footnote{The monopole harmonics $Y_{q,l,m}$ of Ref.
\cite{wu} corresponds to our $\vert k, k_3, k_r\rangle$ with $k=l$,
$k_{3}=m$, $k_{r}=-q$.}

Since the Hamiltonian is made of $J^2$ and $I_r$, we need a complete
set of commuting operators containing them. We can choose
$\vec{J}^2$, $ J_{3}$, $K_{r}$, $I^2$, $I_{r}$ and $\Sigma$. Since
the Gauss law constraint (\ref{norm}) can be written in terms of
these operators as $K_{r}=\Sigma$, the problem reduces to
constructing the simultaneous eigenstates of $\vec{J}^2$, $ J_{3}$,
$I^2$, $I_{r}$, $\Sigma$ from the simultaneous eigenstates of $K^2$,
$K_3$, $I^2$, $I_3$, $\Sigma$. This can be done in general but for
the sake of simplicity we will consider ${1\over2}$-isospin case.
Since $I_r^2={1\over4}$ in this case, the energy spectrum of the
Hamiltonian (\ref{ourham}) is determined by $j$. Other quantum
numbers represent degeneracies. Let $I_3={1\over2} \sigma_3.$ Among
the eigenstates of $I_r$ we choose the ones having $j=0$. They are
given by \beq \vert j=0, i_r=+1/2\rangle=
\left(\begin{array}{c} z_1\\
z_2\end{array}\right),~~~\vert j=0, i_r=-1/2\rangle=
\left(\begin{array}{c} -\bar z_2\\
\bar z_1\end{array}\right).\label{isoeigen}\eeq Now, it can be shown
that the simultaneous eigenfunctions of $\vec{J}^2$, $ J_{3}$,
$I^2$, $I_{r}$ and $\Sigma$ can be obtained by taking the product of
three states \beq \vert j, j_3, i_r, \sigma\rangle=\vert j=0,
i_r\rangle\vert j, j_3, \sigma+i_r\rangle\vert
\sigma\rangle,\label{eig}\eeq where $\vert j, j_3,
\sigma+i_r\rangle$ is the monopole harmonics previously defined. We
have omitted $I^2$ because it is constant.

We illustrate some cases. First, let us consider ${\cal N}=1$ case.
In this case, $\sigma=\pm 1/2$. The ground states are characterized
by $j=0$, for which $j_3=0$ and $j_r=0$. The latter condition
implies that $\sigma+i_r=0$. Consequently, the ground states consist
of the following two states with $E=-1/8$: \bea \vert 0, 0,
-{1\over2}, +{1\over2}\rangle=\vert 0, -{1\over2}\rangle\vert
0,0,0\rangle
\vert+{1\over2}\rangle &=&\left(\begin{array}{c} -\bar z_2\\
\bar z_1\end{array}\right)\vert +{1\over2}\rangle,\nn\\
\vert 0, 0, +{1\over2}, -{1\over2}\rangle=\vert 0,
+{1\over2}\rangle\vert 0,0,0\rangle \vert -{1\over2}\rangle
&=&\left(\begin{array}{c} z_1\\
z_2\end{array}\right)\vert -{1\over2}\rangle.\nn\eea  Note that
$i_r=\sigma=\pm1/2$ states are not allowed. One can check that the
both states are annihilated by $Q$ and $\bar{Q}$. The first exited
states correspond to the value of $j=1$. In this case,
$j_3=(1,0,-1)$,  $i_r=\pm1/2$, $\sigma=\pm1/2.$ Therefore, there are
twelve states with $E=7/8$: \bea \vert 1, (1,0,-1), +{1\over2},
+{1\over2}\rangle &=& \vert 0, +{1\over2}\rangle\vert
1,(1,0,-1),~~1\rangle \vert +{1\over2}\rangle =\bar{z}
\bar{z}\left(\begin{array}{c} z_1\\
z_2\end{array}\right) \vert +{1\over2}\rangle,\nn\\
\vert 1,(1,0,-1), -{1\over2}, +{1\over2}\rangle &=& \vert 0,
-{1\over2}\rangle\vert 1,(1,0,-1),~0\rangle \vert +{1\over2}\rangle
= z \bar z \left(\begin{array}{c} -\bar z_2\\
\bar z_1\end{array}\right)\vert +{1\over2}\rangle,\nn\\
\vert 1, (1,0,-1), +{1\over2}, -{1\over2}\rangle &=& \vert 0,
+{1\over2}\rangle\vert 1,(1,0,-1),~0\rangle \vert -{1\over2}\rangle
=z
\bar{z}\left(\begin{array}{c} z_1\\
z_2\end{array}\right)\vert -{1\over2}\rangle,\nn\\
\vert 1, (1,0,-1), -{1\over2}, -{1\over2}\rangle &=& \vert 0,
-{1\over2}\rangle\vert 1,(1,0,-1),-1\rangle \vert -{1\over2}\rangle
=zz\left(\begin{array}{c} -\bar z_2\\
\bar z_1\end{array}\right) \vert
-{1\over2}\rangle,\label{excited}\eea where the omitted indices on
$zz$, $z\bar{z}$ and $\bar{z}\bar{z}$ are determined depending on
the values of $j_{3}$ and $\sigma+i_{r}$. For instance from
Eq.~(\ref{monopoleharmonics}), one can infer that in the first
line of Eq.~(\ref{excited}), each $j_{3}=(1,0,-1)$ corresponds to
$\bar{z}\bar{z}=(\bar{z}_{1}^2,\bar{z}_{1}\bar{z}_{2},\bar{z}_{2}^2),$
in the second line
$z\bar{z}=(z_2\bar{z}_1,|z_2|^{2}-|z_1|^{2},z_{1}\bar{z}_{2})$ and
in the fourth line $zz=(z_2^2, z_1z_2, z_1^2)$. In general, the
energy eigenfunctions with $j$, $j_3$, $i_r$, $\sigma$ are given
by the states
$$\vert 0, i_r\rangle\vert j, j_3,
\sigma+i_r\rangle\vert \sigma\rangle,$$ with an arbitrary integer
$j$, $j_3$ ($-j\le j_3\le j$), and $i_r$, $\sigma$ satisfying $-j\le
\sigma+i_r\le j$.

Next, let us consider ${\cal N}=2$ case. We have $\sigma =
+1,0,0,-1$, and the corresponding four spin states will be denoted
by $\vert +1\rangle$, $\vert 0_\pm\rangle$, $\vert -1\rangle$. Since
$i_r=\pm1/2$ and $j_r=\sigma+i_r=(3/2, 1/2, -1/2,-3/2)$, $j$ can be
half integral. Also, $j_r$ should satisfy $-j\le j_r\le j$. The
ground states correspond to $j=1/2$ states and are given by \bea
\vert 0,-{1\over2}\rangle\vert {1\over2}, (+{1\over2},-{1\over2}),
+{1\over2}\rangle \vert +1\rangle &=&\bar{z}\left(\begin{array}{c} -\bar z_2\\
\bar z_1\end{array}\right)\vert +1\rangle
,\nonumber\\
\vert 0, +{1\over2}\rangle\vert {1\over2},
(+{1\over2},-{1\over2}),+{1\over2}\rangle\vert 0_\pm \rangle &=&
\bar{z}\left(\begin{array}{c} z_1\\
z_2\end{array}\right)\vert 0_\pm \rangle\nonumber\\
\vert 0, -{1\over2}\rangle\vert{1\over2},
(+{1\over2},-{1\over2}),-{1\over2}\rangle\vert 0_\pm\rangle &=&
z\left(\begin{array}{c} -\bar z_2\\
\bar z_1\end{array}\right)\vert 0_\pm\rangle\nonumber\\
\vert 0, +{1\over2}\rangle\vert {1\over2},
(+{1\over2},-{1\over2}),-{1\over2}\rangle\vert -1\rangle &=&
z\left(\begin{array}{c} z_1\\
z_2\end{array}\right)\vert -1\rangle.\nonumber\eea Altogether, there
are twelve independent ground states with energy $E=1/4$. Again
omitted indices depend on the values of $j_3$. In the first and
second line $j_3=(+1/2,-1/2)$ corresponds to
$(\bar{z}_1,\bar{z}_2)$, and in the last two lines $j_3=(+1/2,-1/2)$
correspond to $(z_2,z_1)$. It can be shown that the above ground
states are invariant under the half of supersymmetry, and
supersymmetry is, in some sense, spontaneously broken from ${\cal
N}=2$ to ${\cal N}=1$.

The analysis can be extended to the general ${\cal N}$. Spin states
can be obtained using the representation \bea
 \beta_{1}&=&1\otimes 1\otimes \cdots 1\otimes
\sigma_-,\nn\\
\beta_{2}&=& 1\otimes 1\otimes \cdots \sigma_-\otimes
\sigma_3,\nn\\
& &\cdots\nonumber\\
\beta_{\cal N}&=&\sigma_-\otimes\sigma_3\otimes\cdots
\otimes\sigma_3, \nn\eea and the eigenvalues of the spin operator is
given by $\sigma=({\cal N}/2, {\cal N}/2-1, \cdots -{\cal N}/2+1,
-{\cal N}/2)$. For ${\cal N}= {\rm odd}$ integer, the ground states
are given by $\vert 0, \mp 1/2\rangle\vert 0,0,0\rangle\vert \pm
1/2\rangle$, and the supersymmetry is unbroken. For ${\cal N}= {\rm
even}$ integer, the ground states are given by $\vert 0, \mp
1/2\rangle\vert 1/2,(+1/2,-1/2),\pm 1/2\rangle \vert \pm 1\rangle$
and $\vert 0, \pm 1/2\rangle\vert 1/2,(+1/2,-1/2),\pm
1/2\rangle\vert 0\rangle$, and half of the supersymmetries is
spontaneously broken.

In summary, we showed that the number of supersymmetries can be
made arbitrarily large for supersymmetric isospin particles on
sphere in the background of specifically chosen spherically
symmetric $SU(2)$ gauge field. The supersymmetry generators form
the standard ${\cal N} =1$ SUSY algebra for a single complex
fermion, $su(2\vert 1)$ algebra for ${\cal N}=2$. But for higher
${\cal N}$, it become the nonlinear realization of the $su({\cal
N}\vert 1)$ algebra. We also gave exact energy spectra and
corresponding eigenfunctions in the case of $I=1/2$ and found that
half of the supersymmetry is spontaneously broken if the complex
number of fermion degrees of freedom is even.  It would be
interesting to investigate details of the nonlinear algebra
Eq.~(\ref{algebra})  and extend the analysis further to general
values of isospin $I$ other than $1/2$.

\acknowledgments We would like to thank Bum-Hoon Lee and Jin-Ho Cho
for useful discussions. The work of STH was supported by the Korea
Research Foundation Grant funded by the Korean Government
(KRF-2006-331-C00071). THL was supported by the Soongsil University
Research Fund. PO was supported by the Science Research Center
Program of the Korea Science and Engineering Foundation through the
Center for Quantum Spacetime(CQUeST) of Sogang University with grant
number R11-2005-021.


\begin{thebibliography}{99}
\bibitem{cole} S. R. Coleman, {\it The Magnetic
Monopole Fifty Years Later} in Les Houches Sum. School (1981) 461;
P. Goddard and D. I. Olive, {\it Rep. Prog. Phys.} {\bf 41} (1978)
1357; R. Jackiw, {\it Ann. Phys.} {\bf 129} (1980) 183.
\bibitem{vine} E. D'Hoker and L. Vinet, {\it Supersymmetry of the Pauli equation
in the presence of a magnetic monopole}, Phys. Lett. B 137 (1984)
72; D. Spector, {\it $N=0$ supersymmetry and the nonrelativistic
monopole}, Phys. Lett. B 474 (2000) 331, hep-th/0001008; M. S.
Plyushchay, {\it On the nature of fermion monopole supersymmetry},
Phys. Lett. B 485 (2000) 187, hep-th/0005122;  C. Leiva and M. S.
Plyushchay, {\it Phys. Lett.} {\bf B 582} (2004) 135,
hep-th/0311150; S. Kim and C. Lee, {\it Supersymmetry-based approach
to quantum particle dynamics on a curved surface with non-zero
magnetic field}, Ann. Phys. 296 (2002) 390, hep-th/0112120.
\bibitem{deJonghe} F. De Jonghe, A. J. Macfarlane, K. Peeters and J. W. van
Holten, {\it New supersymmetry of the monopole}, Phys. Lett. B 359
(1995) 114, hep-th/9507046.
\bibitem{Linden} N. Linden, A.J. Macfarlane and J.W. van Holten,
{\it Particle Motion in a Yang-Mills Field: Wong's Equations and
Spin-${1\over2}$ Analogues}, Phys. Lett. B 373 (1996) 125.
\bibitem{hllo06} S.T. Hong, J. Lee, T.H. Lee and P. Oh, {\it A Complete solution of a
constrained system: SUSY monopole quantum mechanics}, JHEP 0602
(2006) 036, hep-th/0511275; {\it Supersymmetric monopole quantum
mechanics on a sphere}, Phys. Rev. D 72 (2005) 015002,
hep-th/0505018; {\it $N=4$ supersymmeric quantum mechanics with
magnetic monopole}, Phys. Lett. B 628 (2005) 165, hep-th/0507194.
\bibitem{wuya} T.T. Wu and C.N. Yang, in {\it Properties of
matter under unusual conditions}, edited by H. Mark and S.
Fernbach (Interscience, New York, 1969), pp. 349-354; T.T. Wu and
C.N. Yang, {\it Some remarks about unquantized non-Abelian gauge
fields}, Phys. Rev. D 12 (1975) 3843.
\bibitem{freu} A. Pais and V. Rittenberg, J. Math. Phys. {\bf 16} (1975)
2062; P. G. O. Freund and I. Kaplansky, J. Math. Phys. {\bf 17}
(1976) 228; M. Scheunert, W. Nahm and V. Rittenberg, J. Math. Phys.
{\bf 18} (1977) 155; M. Marcu, J. Math. Phys. {\bf 21} (1980) 1277.
\bibitem{bala77} A. P. Balachandran, P. Salomonson, B.-S. Skagerstam and J.-O. Winnberg,
{\it Classical description of a particle interacting with a
non-Abelian gauge field}, Phys. Rev. D 15 (1977) 2308.
\bibitem{wong} S. K. Wong,  {\it Field and particle equations for the classical Yang-Mills
field and particles with isotopic spin},  Nuovo Cim. A 65 (1972)
689.
\bibitem{wu}  T.T. Wu and C.N. Yang, {\it Dirac monopole without
strings: monopole harmonics}, Nucl. Phys. B 107 (1976) 365.

\end{thebibliography}
\end{document}